\documentclass{PoS}
\usepackage{graphicx,multirow,subfigure}
\usepackage{amssymb}
\usepackage{amsmath}
\newcommand{\tabspace}{0.1} 

\title{Supersymmetric Gauged B-L Model of Dark Matter and Fine Tuning}

\ShortTitle{FT/DM in BLSSM}

\author{Luigi Delle Rose\\
        Particle Physics Department, Rutherford Appleton Laboratory, Chilton, Didcot, Oxon OX11 0QX, United Kingdom\\
        E-mail: \email{L.Delle-Rose@soton.ac.uk}}
    
\author{Shaaban Khalil\\
	Center for Fundamental Physics, Zewail City of Science and Technology, Sheikh Zayed,12588 Giza, Egypt\\
	E-mail: \email{Skhalil@zewailcity.edu.eg}}

\author{\speaker{Simon King}\\
	School of Physics and Astronomy, University of Southampton, Highfield, Southampton SO17 1BJ, United Kingdom\\
	E-mail: \email{sjd.king@soton.ac.uk}}

\author{Carlo Marzo\\
	National Institute of Chemical Physics and Biophysics, R{\"a}vala 10, 10143 Tallinn, Estonia\\
	E-mail: \email{Carlo.Marzo@kbfi.ee}}

\author{Stefano Moretti\\
	School of Physics and Astronomy, University of Southampton, Highfield, Southampton SO17 1BJ, United Kingdom\\
	E-mail: \email{S.Moretti@soton.ac.uk}}

\author{Cem S. Un\\
	Department of Physics, Uluda\~{g} University, TR16059 Bursa, Turkey\\
	E-mail: \email{cemsalihun@uludag.edu.tr}}

\abstract{We investigate how the Fine-Tuning (FT) in the B-L Supersymmetric Standard Model (BLSSM) compares to the Minimally Supersymmetric Standard Model (MSSM), where both models have universality. This is done for two scales: both low (i.e. collider) and high (i.e. Grand Unified Theory (GUT)) scales. We see this is similar for both models and the two scale regimes. We also study the possible Dark Matter (DM) candidates each model offers in a realistic scenario satisfying relic density constraints. Our findings are that whilst the parameter space for the single MSSM DM candidate is severely constrained, the BLSSM offers multiple candidates in a much wider region. }

\FullConference{The European Physical Society Conference on High Energy Physics\\
	5-12 July, 2017\\
	Venice}

\begin{document}

\section{Introduction}
One of the most promising theories to address a number of questions left unanswered by the Standard Model (SM) is supersymmetry (SUSY). However, to date, there is no direct experimental evidence supporting any specific BSM model. One of the motivations of SUSY, though not why it was invented, is to solve the hierarchy problem. We know the SM may not be extrapolated to arbitrarily large scales, due the existence of gravity; and the strong hints of unification. So, any states with very large masses will contribute to the self-energy corrections to the Higgs, which would result in a bare mass coupling which is fine-tuned to one part in $\sim 10^{28}$, if the new states were to be at GUT scale, $10^{16}$GeV. As no super-partners have yet been discovered at the LHC, the scale of SUSY is pushed higher and would result in a (much smaller) finely-tuned Higgs mass of order 1 part in $( \Lambda_{SUSY} ^2 /\Lambda_{EW} ^2 )\sim 100$. This FT leads us to think that our first guess of nature obeying the most minimal SUSY scenario, the MSSM, may not have been correct. 

Now, it is certainly worth considering these non-minimal cases and comparing the FT between these and the MSSM. There is no strict absolute definition of FT in nature, but rather it is more of a construct from our point of view. The overall goal of any quantitative measure of FT is to check how different the Universe would be if the fundamental parameters of the theory are changed by an infinitesimal amount. The absolute scale of this is unimportant as it is fundamental to nature, but what is important for our sake is to devise a measure which may be applied to multiple models in order to compare the relative value of their FT. If a given model has particularly large FT, then this may suggest it is unlikely to represent nature. 

In this work, we consider extending the SM gauge group with a $U(1)_{B-L}$, in a SUSY framework. This model includes the benefits of the MSSM, and provides a motivation to add exactly three Right Handed (RH) neutrinos, this in turn allows a natural description of light, non-vanishing, Left Handed (LH) neutrino masses. Another consequence of this model includes several additional DM candidates to the MSSM, which we discuss.

In section \ref{sec:BLSSM} we briefly introduce the BLSSM and its particle content, we then analyse the bounds on this model from colliders and DM searches in section \ref{sec:Collider}. In section \ref{sec:FT} we discuss the FT measures we adopt and in section \ref{sec:Results}, we present all of our findings and then conclude in section \ref{sec:Conclusion}.

\section{The $B-L$ Supersymmetric Standard Model}
\label{sec:BLSSM}
This model extends the SM gauge group to include a gauge symmetry conserving Baryon minus Lepton number ($B-L$). In order to cancel the new triangle anomaly made with three $U(1)_{B-L}$ gauge bosons, one is forced to add three RH singlets to the SM, which we identify as three RH neutrinos. We include a type-I see-saw mechanism such that the larger (TeV scale) mass of the RH neutrinos motivates the small (eV scale) masses of the LH ones. We place this model of an extended gauge group in a SUSY framework \cite{Khalil:2007dr,DelleRose:2017ukx,DelleRose:2017smp}, 
where we may write the superpotential in terms of that of the MSSM with BLSSM specific terms:
\begin{eqnarray*}
		W &=&\mu H_{u}H_{d}+Y_{u}^{ij}Q_{i}H_{u}u^{c}_{j}+Y_{d}^{ij}Q_{i}H_{d}d^{c}_{j}+Y_{e}^{ij}L_{i}H_{d}e^{c}_{j} ~~\big{\}}\text{ MSSM} \nonumber\\
		&+&Y_{\nu}^{ij}L_{i}H_{u}N^{c}_{i} + Y^{ij}_{N}N^{c}_{i}N^{c}_{j}\eta_{1}+\mu^{\prime}\eta_{1}\eta_{2} ~~~~~~~~~~~~~~~~~~~~~\big{\}} \text{ BLSSM-specific}
		\label{superpotential}
\end{eqnarray*}
where the top line is the same as in the MSSM and the bottom shows the BLSSM-specific terms. In the following table we summarise the particle content of the BLSSM.
\begin{table}[h]
	\centering
	\resizebox{0.6\columnwidth}{!}{%
		\begin{tabular}{c | c | c | c | c }
			\multicolumn{2}{c|}{Chiral Superfield} & Spin 0 & Spin 1/2  & $G_{B-L} $\\[\tabspace cm] \hline
			&&&\\[-1em]
			\textcolor{black}{RH Sneutrinos / Neutrinos (x3)} &\textcolor{black}{\(\hat{\nu}\)} & \textcolor{black}{$\tilde{\nu}^* _R$} & \textcolor{black}{$\bar{\nu_R}$}  & \textcolor{black}{(\textbf{1}, \textbf{1}, 0, $\frac{1}{2})$} \\
			
			\textcolor{black}{Bileptons/Bileptinos} &\textcolor{black}{\(\hat{\eta}\)} & \textcolor{black}{$\eta$} & \textcolor{black}{$\tilde{\eta}$}  & \textcolor{black}{(\textbf{1}, \textbf{1}, 0, -1)} \\
			
			&\textcolor{black}{\(\hat{\bar{\eta}}\)} & \textcolor{black}{$\bar{\eta}$} & \textcolor{black}{$\tilde{\bar{\eta}}$}  & \textcolor{black}{(\textbf{1}, \textbf{1}, 0, 1)} \\[0.5em] \hline
			
			\multicolumn{2}{c|}{\phantom{\Large{l}} Vector Superfields \phantom{\huge{l}} } & Spin 1/2 & Spin 1 & $G_{B-L} $\\ \hline	
			\multicolumn{2}{c|}{\color{black} \phantom{\huge{l}} BLino / B' boson \phantom{\huge{l}} } & \color{black} $\tilde{B}^{\prime 0}$ & \color{black} $B^{\prime 0}$ & \color{black}(\color{black}\textbf{1} \color{black}\textbf{1}, \color{black}0, \color{black}{0)}		
		\end{tabular}%
	}
\end{table}

In addition to the MSSM, we have the three RH neutrino states and their superpartners the ``sneutrinos''. Two $B-L$ complex Higgs singlets, the ``bileptons'' to break the $U(1)_{B-L}$ (two because of the anomaly cancellation condition, as with the Higgs doublets in the MSSM), and their superpartners, the ``bileptinos''. Finally we have a new B' gauge boson associated with this group and its superpartner the ``BL-ino''. Both the sneutrinos and BL-ino/bileptino-like neutralinos make good DM candidates, which we investigate later. 



\section{Collider and DM bounds}
\label{sec:Collider}
We compare the MSSM and BLSSM both with complete universality, ie $g_1 = g_2 = g_3 (= g_{BL} )$ at GUT scale. The \texttt{SARAH} and \texttt{SPheno} programs \cite{Staub:2013tta,Porod:2003um} use the high scale parameters as initial inputs and calculate the low-scale spectra. In this work we have chosen to fix the mass of our $Z'$, rather than generating a spectra of low mass unphysical points. We require that this mass satisfies Electro-Weak Precision Observables (EWPOs) from both LEP2 data and the LHC with $\sqrt{s}=13$ TeV and an integrated luminosity of $\mathcal{L}=13.3$ fb$^{-1}$ as presented in \cite{Accomando:2016sge}. A conservative estimate of $M_{Z'}=4$ TeV allows a realistic mass given our particular model for all couplings and widths. We also enforce that our heavy scalars, such as those found in a 2HDM \cite{Arhrib:2016rlj}, plus our $B-L$ Higgs are allowed by LHC searches, by using the \texttt{HiggsBounds/HiggsSignals} programs \cite{Bechtle:2008jh,Bechtle:2011sb,Bechtle:2013wla,Bechtle:2015pma,Bechtle:2013xfa}, this removes scalars which are excluded by current bounds and also demands the lightest CP-even scalar must have SM-like couplings. To check the relic density for each of our spectrum points, we use the \texttt{MicrOMEGAs} program \cite{Belanger:2006is,Belanger:2013oya}. We compare this calculated value to the current measured value of the DM relic density:
\begin{equation} \label{PLANCK}
\Omega h^2 = 0.1187 \pm 0.0017({\rm stat}) \pm 0.0120({\rm syst}) 
\end{equation}
as measured in 2015, by the Planck Collaboration \cite{Ade:2015xua}.

\section{Fine-Tuning Measures}
\label{sec:FT}
There exist many quantitative measures of FT, all of which numerically estimate how resistant a given model is to some shift in the initial parameters. The most popular of these measures the change in mass of the SM Z boson mass at EW level when shifting one of the fundamental parameters of the theory by some infinitesimal amount \cite{Ellis:1986yg,Barbieri:1987fn},
\begin{equation}
\Delta={\rm Max} \left| \frac{\partial \ln v^2}{\partial \ln a_i}\right| =  {\rm Max} \left| \frac{a_i}{v^2} \frac{\partial v^2}{\partial a_i } \right|  = {\rm Max} \left| \frac{a_i}{M_Z ^2} \frac{\partial M_Z ^2}{\partial a_i} \right|.
\label{eq:BGFT}
\end{equation}
In this work we check the FT at two scales, at both high scale (the common literature scale) and low-scale. The reason to perform this check at two scales is to check that a model does not have a low fine-tuning at GUT scale but be very finely tuned at EW. To calculate the FT at GUT scale, we use the fundamental parameters: the unification masses for scalars ($m_0$)  and gauginos ($m_{1/2}$), the universal trilinear coupling ($A_0$), the Higgsino mass $\mu$ and quadratic soft SUSY term ($B\mu$), for the MSSM. These are also applicable to our BLSSM scenario, but with further terms: the bileptino mass $\mu '$ and quadratic term $B\mu '$. The parameters are thus:
\begin{equation}
a_i = \left\lbrace m_0 , ~ m_{1/2},~ A_0,~ \mu ,~ B \mu ,(~\mu' ,~B\mu ') \right\rbrace.
\end{equation}
Whilst it is normally viewed that loop corrections do not affect the FT, recent work \cite{Ross:2017kjc} has shown that they can change the absolute scale by a factor of two. We now turn to the SUSY-scale FT. For the BLSSM (noting one obtains the MSSM by taking $X \rightarrow 1$, $\tilde{g} \rightarrow 0$), one may minimise the Higgs potential and solve the tadpole equations to find the mass of the SM Z in terms of SUSY-scale quantities,
\begin{equation}
\frac{Mz^2}{2}=\frac{1}{X}\left( \frac{ m_{H_d}^2 + \Sigma _{d} }{ \left(\tan ^2(\beta
	)-1\right)}-\frac{ (m_{H_u}^2 + \Sigma _u) \tan ^2(\beta )}{
	\left(\tan ^2(\beta )-1\right)} + \frac{\tilde{g} M_{Z'}^2 Y}{4 g_{BL}
}- \mu ^2 \right), \label{eq:blssm_mz}
\end{equation}
where
\begin{align}
X&= 1 + \frac{\tilde{g}^{2}}{(g_{1}^{2}+g_{2}^{2})}+\frac{\tilde{g}^{3}Y}{2g_{BL}(g_{1}^{2}+g_{2}^{2})}, \\
Y&= \frac{\cos(2\beta ')}{\cos (2\beta)} = \frac{\left(\tan^2 {\beta} +1\right) \left(1-\tan^2 {\beta '} \right)}{\left(1-\tan ^2 {\beta } \right) \left(\tan ^2 {\beta '},
	+1\right) },
\end{align}
and
\begin{equation}
\Sigma_{u,d}= \frac{\partial \Delta V}{\partial v_{u,d} ^2}
\end{equation}
are the loop corrections. If one treats these as independent contributions, as done in \cite{Baer:2012up}, then since the FT measure is only the maximum contribution, then these small terms do not affect the FT. Note that this is a different treatment than in the GUT-FT case, but it is the comparison of two models within a given scheme which is of importance to us. The parameters analogous to the high-scale are 
\begin{equation}
\Delta_{\rm SUSY}\equiv {\rm Max}(C_{i})/(M_{Z}^{2}/2)~,
\label{FT}
\end{equation}
where our FT parameters are $C_i$, which represent each of the terms in \ref{eq:blssm_mz}. One may guess a large $M_{Z'}$ dominates and so the BLSSM would have a larger FT than the MSSM, but the low value of the gauge-kinetic mixing $\tilde{g}$ suppresses this. What dominates, in all four scenarios of low/high scale for MSSM/BLSSM, is the $\mu$ term. Other parameters offer negligible contributions.

\section{Results}
\label{sec:Results}
We scan over the parameter space of the MSSM/BLSSM where each model has 60,000 data points which satisfy the \texttt{HiggsBounds/HiggsSignals} requirements. For each point we calculate a full spectrum, including the LSP DM candidate relic density and a value for low- and high-scale FT. Our scan ranges for both models are over the regions: [0,5] TeV for $m_0$ and $m_{1/2}$, [0,60] for $\tan \beta$, [-15,15] TeV for $A_0$ and, for the BLSSM alone, [0,2] for $\tan \beta '$ and $[0,1]$ for the neutrino Yukawa couplings $Y^{(1,1)}$, $Y^{(2,2)}$, $Y^{(3,3)}$, with $M_{Z'}=4$TeV.

\begin{figure}[ht!]
	\newcommand{\size}{0.50}
	\subfigure[BLSSM GUT-FT.]{\includegraphics[scale=\size]{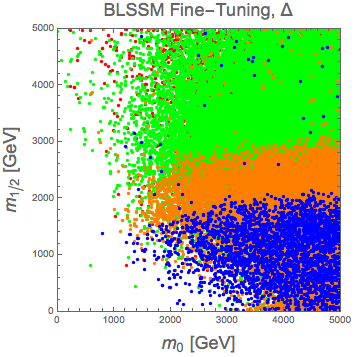} \label{fig:ftbg_blssm_m0_m12}}	
	\subfigure[BLSSM SUSY-FT. ]{\includegraphics[scale=\size]{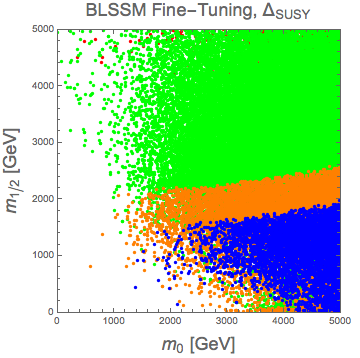}\label{fig:ftew_blssm_m0_m12}}

	\subfigure[MSSM GUT-FT.]{\includegraphics[scale=\size]{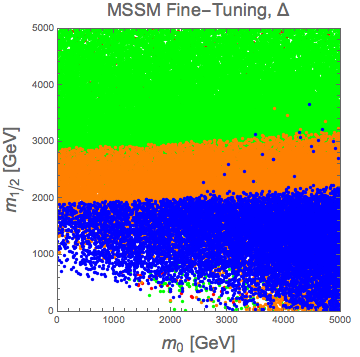}\label{fig:ftbg_mssm_m0_m12}}
	\subfigure[MSSM SUSY-FT. ]{\includegraphics[scale=\size]{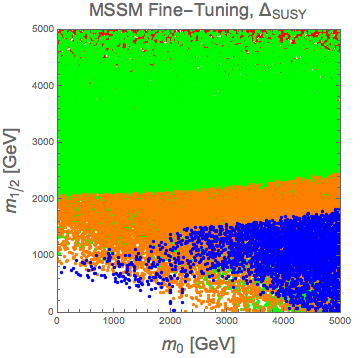}\label{fig:ftew_mssm_m0_m12}}
	\caption{Fine-tuning in the plane of unification of scalar, gaugino masses for BLSSM and MSSM for both GUT-parameters ($\Delta$) and SUSY parameters ($\Delta_{\rm SUSY}$). The FT is indicated by the colour of the dots: blue for FT $<$ 500; Orange for 500 $<$ FT $<$ 1000; Green for 1000 $<$ FT $<$ 5000; and Red for FT $>$ 5000.}
	\label{fig:all_m0_m12}
\end{figure}

Figure \ref{fig:all_m0_m12} shows the FT for all four cases - both high and low-scale FT for the BLSSM and MSSM. We plot the FT in the $m_0$-$m_{1/2}$ plane, rather than any of the other initial parameters ($\tan \beta$, $A_0$, and for the BLSSM, $\tan \beta '$), recalling that others, eg $\mu$, are determined by these. The FT is coloured for each point: red for FT $>$ 5000, green for 1000 $<$ FT $<$ 5000, orange for 500 $<$ FT $<$ 1000 and blue (the least finely-tuned points) for FT $<$ 500. We immediately see that the FT is similar between the BLSSM and MSSM. Both are approximately independent of $m_0$ and have a strong dependence on $m_{1/2}$, which largely determines $\mu$, which dominates the FT. The lack of points at $m_0 < 1000$GeV for the BLSSM is due to the $M_{Z'}=4$TeV requirement. A more detailed analysis of these results may be found in \cite{DelleRose:2017smp}.

\begin{figure}[h!]
	\centering
	\includegraphics[scale=0.4]{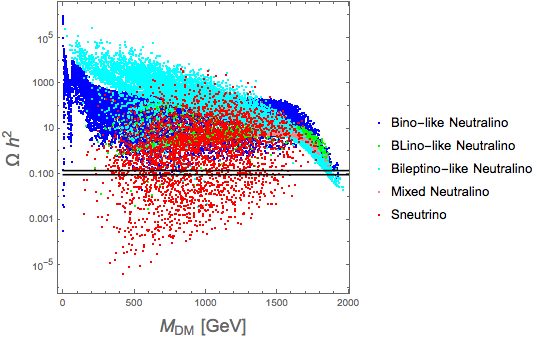}
	\includegraphics[scale=0.4]{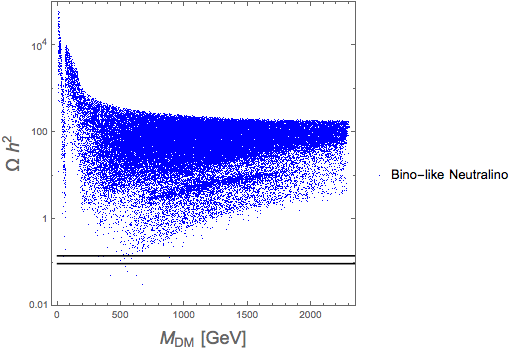}
	\caption{(a) Relic density vs LSP mass for the BLSSM.   
		(b) Relic density vs LSP mass for the MSSM. In both plots the horizontal lines identify the $2\sigma$ region around the current central value of  $\Omega h^2$. }
	\label{fig:BLSSMvsMSSM-DM}	
\end{figure}

Now we consider the DM sectors of both models. In the universal MSSM, only one good candidate, a Bino-like neutralino, survives. For the BLSSM, this candidate exists, but we have additional neutralinos from the SUSY partners to the B' boson and new Higgses: the BL-ino and Bileptino-like neutralinos. Finally, we have the SUSY partner to the TeV scale RH neutrinos, the sneutrinos. We plot the relic density vs mass for these candidates in figure \ref{fig:BLSSMvsMSSM-DM}. Immediately one sees for the MSSM, that only $\sim$5 of the 60,000 spectrum points survive the relic constraint. For the BLSSM, one finds a more promising bino-like neutralino sector (many more of these blue points satisfy the relic constraints), but, each of the other types of neutralinos have points which satisfy the relic requirements too. The most promising candidate, in terms of allowed parameter space, is the sneutrino, whose points are in red. The relic density requirements clearly pass directly through the middle of these points, allowing many viable DM candidates.

\section{Conclusions}
\label{sec:Conclusion}
We have examined the FT and DM properties of the universal scenarios of the MSSM and BLSSM. We see that whilst the FT is similar at both low and high scales for the two models, when including a DM candidate which satisfies the relic density limits, the MSSM is very highly constrained, whereas the BLSSM allows large regions of parameter space.

\section*{Acknowledgements}
\noindent
SM is supported in part through the NExT Institute. The work of LDR has been supported by the ``Angelo Della Riccia'' foundation and the STFC/COFUND Rutherford International Fellowship scheme. The work of CM is supported by the ``Angelo Della Riccia'' foundation and by the Centre of Excellence project No TK133 ``Dark Side of the Universe''. The work of SK is partially supported by the STDF project 13858. All authors acknowledge support from the grant H2020-MSCA-RISE-2014 n. 645722 (NonMinimalHiggs).

\end{document}